# Phonon anomalies and possible local lattice distortions in giant magnetocapacitive $CdCr_2S_4$

V. Gnezdilov[1], P. Lemmens[2], Yu. G. Pashkevich[3], C. Payen[4], K. Y. Choi[5], J. Hemberger[6], A. Loidl[6], V. Tsurkan[6,7]

[1] B. I. Verkin Inst. for Low Temperature Physics, NASU, 61164 Kharkov, Ukraine;
[2] Inst. for Condensed Matter Physics, TU Braunschweig, D-38106 Braunschweig, Germany;
[3] A. A. Galkin Donetsk Phystech NASU, 83114 Donetsk, Ukraine;
[4] Institut des Materiaux Jean Rouxel, CNRS-Universite de Nantes, BP32229, 44322 Nantes cedex 3, France;
[5] Department of Physics, Chung-Ang University, 221 Huksuk-Dong, Dongjak-Gu, Seoul 156-756, Republic of Korea;
[6] Experimental Physics V, Center for Electronic Correlations and Magnetism, University of Augsburg, D-86135 Augsburg, Germany;
[7] Inst. of Applied Physics, Academy of Sciences of Moldova, MD-2028 Chisinau, R. Moldova

(May 8th, 2011)

We investigate the origin of large magnetocapacitive effects in $CdCr_2S_4$ single crystals focusing on phonon anomalies in the fluctuation regime $T<T^*=130K$ and with the onset of ferromagnetic order, $T_c$=84.4 K. The observation of longitudinal optical and formerly infrared active modes for $T<T_c$ is used to propose a loss of inversion symmetry and Cr off-centering. Further anomalies in intensity, frequency, and linewidth have an onset temperature $T^*$, that coincides with the onset of magnetocapacitive effects. The respective intensity anomalies are attributed by part to an enhanced electronic polarizability of displacements that modulate the Cr-S distance and respective hybridization. This is taken as a prerequisite to the proposed Cr off-centering. Anomalies due to the previously disregarded photodoping and resonance are analyzed comparing different excitation energies.

# I. INTRODUCTION

The coexistence and mutual coupling of magnetic and dielectric properties in solids are presently discussed from both fundamental and application point of view[1]. Large coupling effects have been found in different material classes, ranging from antiferromagnets (AF) with spiral ground states[1], via composite or lamellar materials that provide magneto-elastic coupling at interfaces[2] to lone pair systems with hybridized electronic states that gain energy by polar distortions[3]. The intrinsic coexistence, however, of ferromagnetic and ferroelectric order parameters is hindered by exclusive symmetry conditions and having occupied 3d electron states to provide magnetic exchange and their energetic disadvantage in polar distorted coordinations[4].

Among the reported materials, the cubic thio-spinel $CdCr_2S_4$ stands out due to its large ferromagnetic (FM) moment ($T_c$ = 84.4 K) coexistent with a remanent dielectric polarization. Furthermore, a pronounced fluctuation regime with relaxor ferroelectric behavior[5] and evidence for polar nanodomains exists above $T_c$. This is evident from a maximum of the dielectric loss at $T_{max}(\omega \rightarrow 0) = T^* = 130$ K and a broadening of X-ray peaks[6]. Most unusual in these observations is the $Fd\bar{3}m$ space group symmetry of the paramagnetic phase. This group symmetry strictly forbids the existence of a ferroelectric order parameter. Therefore, mesoscopic or microscopic variations of the electronic state which create local distortions may play an important role for the relaxation effects and magnetocapacitive effects[7]. In its turn magnetocapacity above $T_c$ might arise due to magnetoelectric coupling between magnetic moment fluctuations and local distortions. Similar to manganites the cross susceptibilities could be magnified by microscopic phase separation[8,9]. As suggested by Hemberger et al.[5] off-centering of Cr-ions in $CdCr_2S_4$ could serve as an intrinsic mechanism for this effect. Also in far–infrared studies[10] of $CdCr_2S_4$ significant effects are detected in the temperature dependence of the plasma frequencies, indicating changes in the nature of the bonds and significant charge redistribution.

On the other hand there is evidence for a nonintrinsic nature of some dielectric anomalies (see discussion in Ref.[9,10]). Problems may arise due to non-stoichiometry, finite conductivity and possible internal interfaces of some samples that impede a proper dielectric characterization. Nevertheless, record values of magnetocapacitive effects with an increase of the dielectric constant $\varepsilon'$ of up to a factor of 30 in a magnetic field of 10 T are reported[7]. Due to the lack of microscopic information the understanding of the underlying physics is rather limited and requires further spectroscopic investigations as well as high quality samples prepared by different growth methods.

In the present paper, the observation of pronounced anomalies of the phonon systems based on Raman light scattering (RS) on single crystals of $CdCr_2S_4$ is reported. Following an intense study and optimization of preparation parameters we use $CdCr_2S_4$ single crystals grown by Br transport. These crystals are expected to be free from transport agent contaminations. However they are smaller in size than samples from Cl transport. Based on the observation of longitudinal polar phonons in our Raman spectra we suggest a non-centrosymmetric $F\bar{4}3m$ crystal structure at low temperatures (T<$T_c$) clearing up a long standing dispute[5,11] on a possible Cr off-centering and loss of inversion symmetry. Polar phonons that modulate the Cr-S distances show large intensity gains in magnetically ordered phase that point to an intrinsic magnetoelectric coupling effect of the magnetic degrees of freedom and local polarizations. Furthermore, we demonstrate that the onset of these anomalies exist already in the fluctuation region, for $T_c < T < T^*$.

## II. EXPERIMENT

RS experiments were performed on as-grown, shiny [111] surfaces of $CdCr_2S_4$ single crystals prepared from a Br transport reaction. Spectra of the scattered radiation were collected by a Dilor-XY triple spectrometer and recorded by a nitrogen-cooled charge-coupled device detector. We used the excitation wavelengths λ = 632.8 nm of a HeNe laser, λ = 488 nm of an Ar/Kr ion laser, and λ = 532.1 nm of a solid-state laser with a power level P < 6 mW, i.e. a factor 15 smaller than in a previous studies[12-14]. In our experiments we used parallel (*XX*) and crossed (*XY*) light polarizations. Temperature dependencies were investigated using a variable temperature closed cycle cryostat (Oxford/Cryomech Optistat, RT-2.8 K) and a continuous helium flow optical cryostat (Cryovac, RT-3K).

## III. RESULTS AND DISCUSSION

Earlier RS experiments[12-14] showed agreement with factor group calculations for the room temperature structure $Fd\bar{3}m$ which is cubic and inversion symmetric[15]. The decomposition of the Raman-active symmetry components leads to $\Gamma = A_{1g}(394) + E_{1g}(257) + 3T_{2g}(99, 280, 352)$ with the calculated phonon frequencies given in brackets (cm$^{-1}$)[15,16].

In Fig. 1 Raman spectra of $CdCr_2S_4$ measured in two scattering geometries are presented. At room temperature very sharp maxima are observed thus proving the high quality of the used single crystals. Also the background level of the spectra is uniform and flat. Samples that are defect rich or non-stoichiometric show broad maxima due to fluorescence effects. The number and frequencies of the excitations are in detailed agreement with lattice calculations[16] based on the space group $Fd\bar{3}m$. The modes are marked by dashed lines and with the respective symmetry representation. The $A_{1g}$ phonon line is observed only in *XX* scattering geometry without any leakage in the *XY* scattering geometry. With decreasing temperature several anomalies develop. Most spectacular is the appearance of a new mode at 106 cm$^{-1}$, i.e. 8 cm$^{-1}$ higher than the lowest frequency mode at 98 cm$^{-1}$. At the same time, the high frequency phonons with $\Delta\omega > 250$ cm$^{-1}$ show different intensity versus temperature dependencies for different incoming photon energies. While all four phonon modes show an enormous gain in intensity with the excitation wavelengths $\lambda = 532.1$ nm, only two phonon modes with $\Delta\omega > 300$ cm$^{-1}$ show a similar behavior with the excitation wavelengths $\lambda = 632.8$ nm.

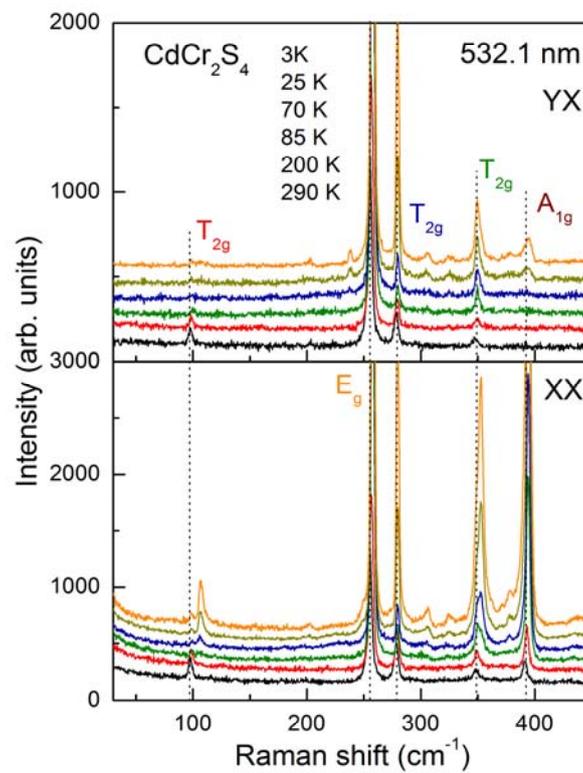

FIG. 1. RS spectra of CdCr$_2$S$_4$ as function of temperature taken with $\lambda = 532.1$ nm. The curves are shifted for clarity. The phonon lines are marked by dashed lines and assigned according to the high temperature structure[15].

In proximity of the 352-cm$^{-1}$ $T_{2g}$ mode and the 396-cm$^{-1}$ $A_{1g}$ mode the evolution of sidebands is observed at low temperatures. The $T_{2g}$ mode at 280 cm$^{-1}$ does not split, see Fig. 2. The temperature dependence of the peak frequencies shows a change of slope at *T\** and sidebands appear for *T*<*T*$_c$. The splittings themselves are rather small and of the order of 1-2% for the high frequency modes. These observations are not compatible with a large reduction of symmetry and point to a preserved cubic symmetry.

In the same figure the approximate frequencies of the low-temperature longitudinal optical (LO) modes from an earlier infrared absorption experiment are given by arrows[10,17]. It is obvious that the observed new modes are very close or coincide with these data within the accuracy of the earlier Kramers-Kronig analysis. For the 280-cm$^{-1}$ $T_{2g}$ mode no corresponding LO phonon exists[18] and therefore no splitting can be detected. With a loss of inversion only LO phonons, formerly IR active modes, develop a RS cross section in back scattering geometry.

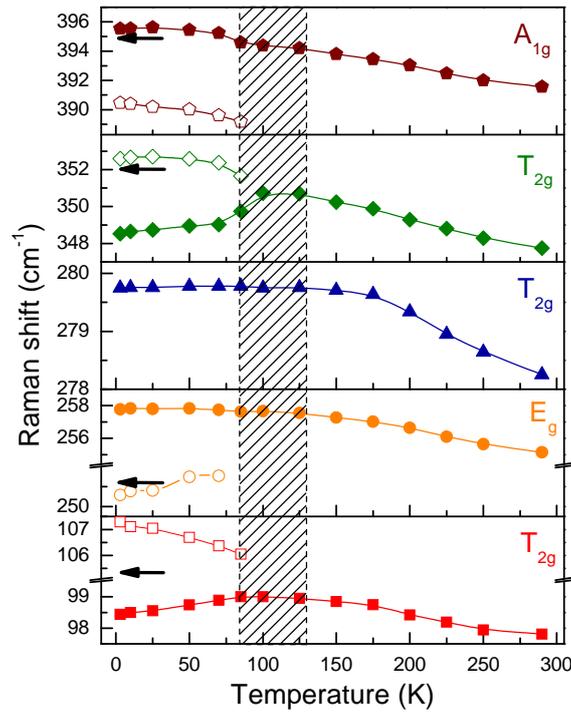

FIG. 2. RS phonon frequency of CdCr$_2$S$_4$ as function of temperature. The fluctuation regime between $T_c$ = 84.4 K and $T^*$ = 130 K is marked by a dashed bar. Arrows mark the approximate frequency of LO phonons at $T$ = 10 K from earlier IR absorption experiments[17].

In Fig. 3 the phonon intensities of the high temperature Raman active modes and additional modes are given. Both characteristic temperatures are evident in the data. Noteworthy is the large enhancement of the $E_g$ mode and high frequency $T_{2g}$ and $A_{1g}$ modes in the fluctuation regime between $T_c$ and $T^*$ at excitation under λ = 532.1 nm. It leads to an overall increase by a factor 5 ($E_g$) and 25 ($T_{2g}$, $A_{1g}$) to low temperatures. Interestingly, the intensity of the LO phonons increases with a similar slope down to lowest temperatures. The intensity of the 280-cm$^{-1}$ mode is nearly constant between $T_c$ and $T^*$ and enhances by a factor 8 to low temperatures. The intensity of the 98-cm$^{-1}$ mode drops in the whole temperature region with a sharp slope changing at $T_c$. In contrast the intensity of the 280-cm$^{-1}$ and the 255-cm$^{-1}$ modes both drop by 70% in the high temperature regime and recover only to about 1/3 of this intensity at low temperatures with an excitation at $\lambda = 632.8$ nm.

The RS intensity of a phonon mode is proportional to the scattering volume and to the square of the electronic polarizability with respect to the displacement of the particular mode. A change of the surface conductivity may indeed obscure the intrinsic intensity evolution of phonon modes. However, the scattering volume and optical penetration depth develop in parallel for all modes. Therefore, the observed, very specific anomalies must at least be partly related to the local enhancement of electronic polarizability.

In Fig. 4 we show the temperature dependent linewidths (full width at half maximum, FWHM) of the phonon modes in $CdCr_2S_4$. The $T$ dependencies of the FWHM of all modes show anomalies in the fluctuation regime and at the magnetic ordering temperature.

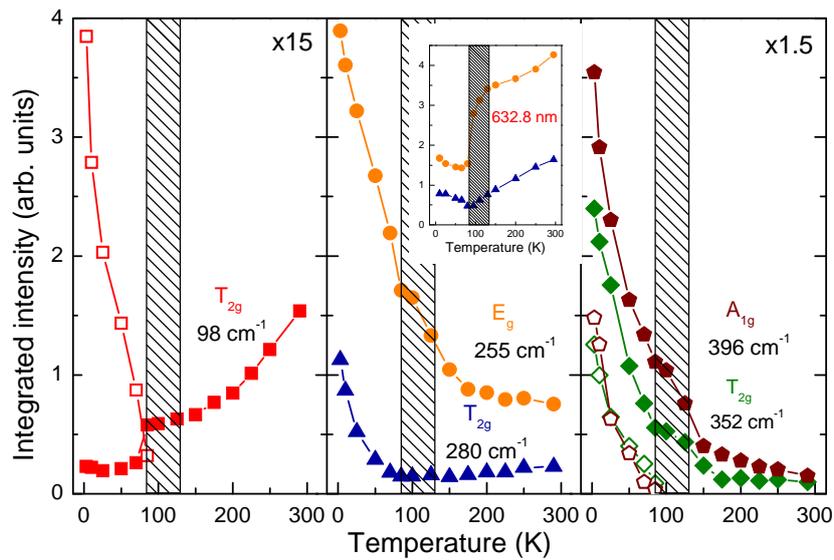

FIG. 3. Integrated RS intensity of $CdCr_2S_4$ as function of temperature for the high energy and the low energy modes in the right panel (an exciting laser wavelength of $\lambda = 532.1$ nm). The open symbols in the left right panels correspond to the low temperature modes at 106, 352 and 390 cm$^{-1}$, respectively. The inset in the middle panel shows the results of the RS experiment with $\lambda = 632.8$ nm.

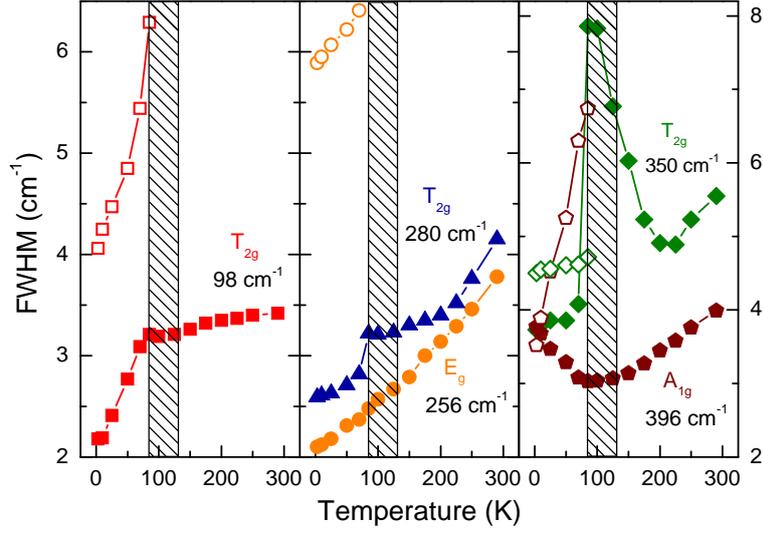

FIG. 4. The Raman phonon widths as a function of temperature of $CdCr_2S_4$.

The left panel of Fig. 5 shows the low energy $T_{2g}$ and the related LO mode close to $T_c$. These modes demonstrate the largest splitting which facilitates their discrimination. With the onset of FM order the $T_{2g}$ mode is damped, The LO appears clearly at the magnetic transition. This behavior is different from the intensity gain that the high energy $T_{2g}$ mode shows for temperatures around $T^*$. At lowest temperature the intensity of the sideband is comparable to the high temperature intensity of the $T_{2g}$ mode.

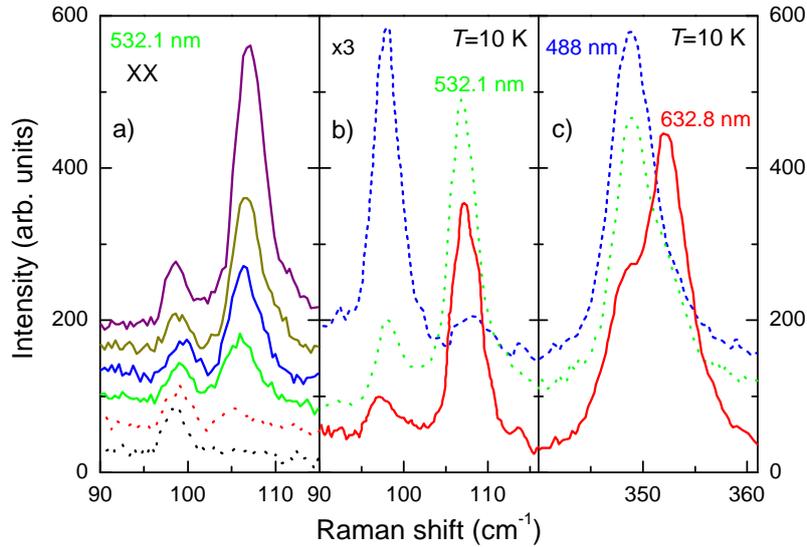

FIG. 5. a) Evolution of the low frequency $T_{2g}$ mode with an excitation wavelength of $\lambda = 532.1$ nm at $T = 10, 25, 65, 80, 95, 110$ K, from top to bottom. b) and c) RS experiments with $\lambda = 488$ nm (top, dashed line), 532.1 nm (middle, dotted line), and 632.8 nm (bottom) at $T = 10$ K. Curves have been shifted for clarity.

Previous investigations using circular polarized light and optical pumping showed that resonance effects of the excitation wavelength with electronic states in $CdCr_2S_4$ have to be considered[13]. These effects certainly complicate the interpretation of Raman scattering experiments. On the other side they open up the opportunity to modify the local electronic properties of $CdCr_2S_4$[19] by an internal photo effect and an electron transfer from the highly structured valence band of hybridized Cr $t_{2g}$ -S $2_p$ states to an unoccupied Cr $e_g$ - Cd s high energy continuum[16]. The corresponding gap is $\Delta$ = 1.6-1.8 eV [20].

In the following we will show evidence for a strong coupling of the phonon system to exactly these states. In Figs. 5b and c we compare RS with the previously used excitation wavelength $\Delta$ =632.8 nm (1.97 eV) and 488 nm (2.55 eV), respectively. The difference is rather drastic. A reversal of the effect of temperature on the RS spectra is evident, i.e. the intensity ratio of the $T_{2g}$ to the LO phonon with 488 nm is close to that at $T \approx T_c$ with 632.8 nm. This effect cannot be attributed to a modification of the selection rules similar to observations in resonance scattering on semiconductors[21] or a varying penetration depth as the intensities of both modes are affected. We attribute it to a local occupation of the antibonding Cr $e_g$ - Cd s states within the conduction band. The high selectivity of the Raman intensity to the laser excitation energy reflects the difference in the lattice dynamics in which different phonons modulate different electronic bonds. Note, that in accordance with the calculations by Lutz et al. [15] the vibration of 95 cm-1 $T_{2g}$ phonon mainly involves Cd ions displacements, while to the LO 105 cm-1 phonon, besides Cd vibrations, the Cr and S ions displacements also contribute.

Understanding the anomalous magnetocapacitive behavior of $CdCr_2S_4$ involves two issues. The first addresses the inconsistency between the high temperature structure and the observed properties, i.e. global or local deviations from inversion symmetry. The second question concerns the interplay between local distortions and the spin polarization, i.e. the magnetocapacitive coupling mechanism. It has to be outlined that a nonlinear relation between the corresponding fields and susceptibilities that is compatible with the observed phonon anomalies is a perquisite but no proof of true magnetoelectricity[3,4]. In addition spontaneous moments of proper magnitude should develop with decreasing temperatures.

With the present RS data we give evidence that in $CdCr_2S_4$ the inversion symmetry is lost and the resulting low temperature symmetry is reduced from $F d\bar{3}m$ to $F\bar{4}3m$. As shown in Fig. 6 a frozen $A_{2u}$ mode corresponds to the long suggested Cr-off centering[16]. The preserved cubic symmetry of $F\bar{4}3m$ is the only way to describe the observed excitation spectrum[22]. Note, that electric polarization is also not allowed by the symmetry of $F\bar{4}3m$ group. However, static $A_{2u}$ type distortions provide magnetoelectric coupling between magnetic and polar degrees of freedom. This low temperature structure allows invariants of the free energy F $\propto$

$M_xM_yP_z+M_yM_zP_x+M_xM_zP_y$ with an electric polarization $P$ with some direction to a net magnetic moment $M$ [22]. Such kind of invariant was exactly prohibited in the earlier accepted $Fd\bar{3}m$ space group.

Cubic symmetry is conserved by $A_{2u}$ with Cr and S static displacements, two nonequivalent Cr sites and two alternating S-Cd bond lengths. In the new symmetry the atom occupied sites are 4a-Cd(1), 4d-Cd(2), 16e-Cr, 16e-S(1), 16e-S(2). The additional modes are a result of the related $A_{2u}$ displacement that evolves into a long range distortion for $T<T_c$. The number of Raman allowed modes are $\Gamma = 3A_1 + 3E + 7T_2$. Besides the previously observed $1A_{1g}(A_1)$, $1E_g(E)$ and $3T_{2g}(T_2)$ this number is enlarged by the former $2A_{2u}(A1)$ and $2E_u(E)$ silent modes related to S- and Cr-displacements as well by $4T_{1u}(T_2)$ of far infrared polar modes. Calculations give $A_{2u}$ modes[23] at 331 and 422 cm$^{-1}$ and $E_u$ modes 214 and 357 cm$^{-1}$. Those calculations have been done on the basis of semi empirical models. Nevertheless, below $T_c$ we observe in the respective energy range additional lines at 238, 371 and 378 cm$^{-1}$ which could be attributed to the previously silent modes.

Beside the single-phonon scattering, some lines in our Raman spectra of $CdCr_2S_4$ can be interpreted as a multiple-order scattering that is a finger print of lattice anharmonicity. Note, that assuming a second order phase transition from $Fd\bar{3}m$ to $F\bar{4}3m$ in the high temperature phase one of $A_{2u}$ modes must demonstrate a soft mode behavior. However, this cannot be detected neither by far-infrared nor by Raman studies. In its turn, under approaching of $T_c$ from low temperature phase one of $A_1$ modes should be a soft mode and can be seen in Raman spectra.

The intensity/polarizability enhancements of all high frequency modes are strong evidence for pronounced non - linearities of the coupled electronic-lattice systems. The intensity of these modes and the LO sidebands is still increasing in the temperature range where the macroscopic magnetization does not change so much[5]. In Fig. 6 we sketch displacements with enhanced polarizability and it is evident that all modes effectively modulate the hybridization of the Cr $t_{2g}$ - S 2p states by a simultaneous stretching - squeezing of the S-Cd and S-Cr bonds, respectively[12,15]. In contrast the intermediate frequency modes show an antiphase stretching on the opposite sides of the S-Cr cube. Depending on the relative amplitude of the Cr and S displacements such a motion can cancel the contribution of all S-Cr bonds to the dielectric response as well as to the RS response. This is to our opinion the origin of the minimum observed in the intensity of, e.g. the 255-cm$^{-1}$ mode with 632.8nm excitation and for temperatures below $T_c$. The displacements sketched in Fig. 6 could also be interpreted as a Cr-Cr dimer formation. Such structural correlations lead to large energy gains on frustrated topologies, i.e. it leads to the spin gap formation in $MgTi_2O_4$[24] and $CuIr_2S_4$[25]. For $CdCr_2S_4$ the hybridized states including some nonstoichiometry provide a finite but small density of states at the Fermi energy[26].

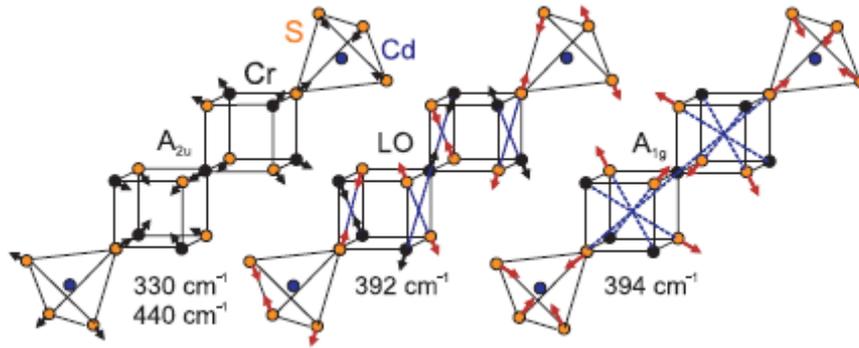

FIG. 6. Displacement patterns and excitation energies of different modes that lead to a large modulation of the Cr-S hybridization[15,23].

The presently discussed magnetoelectrics provide a rather complex picture of possible mechanisms for large cross susceptibilities[1]. Systematic investigations of the phonon frequencies and the magnetic ground state of different chromium spinels show that $CdCr_2S_4$ is close to a bond frustrated phase with enhanced spin-electron-phonon coupling[10]. For some rare earth perovskites $RCrO_3$ similar conclusions have been drawn and the observation of multiferroicity attributed to certain modulations of the Cr-O-Cr bond angle and length[27]. To describe the tendency for spontaneous polarizations and off-centering[28,29] the concept of a second order Jahn-Teller effect has been used. This effect is based on the hybridization of nearly degenerate electronic states of appropriate symmetry[3,4] that overcompensates the cost of Coulomb energy due to the distortions of occupied 3d electron systems. In multiferroic $BiFeO_3$ the off-centering of Bi is a result of a hybridization of both the Bi $6s^2$ and unoccupied $6p^0$ states with O 2p orbitals leading to a highly polarizable lone pair on one side of the Bi ion[30]. Weaker displacements and anomalies of the Raman scattering intensity have been observed for the case of $Cd^{2+}$ ions in sulphur environments, e.g. in $CdPS_3$[31]. We therefore attribute the enhanced local polarizabilities caused by diagonal S displacements in $CdCr_2S_4$ to hybridization effects of moderately correlated Cd and S states [16]. This also agrees to recent observations of the effect of electric fields on electric transport [32].

The lattice degrees of freedom provide another peculiarity. The energy splitting between polar and nonpolar modes in $CdCr_2S_4$ is extremely small compared to other energy scales of the system. Evidently, it reflects some internal degeneracy of the lattice subsystem. This accidental degeneracy can easily be removed by third or higher order anharmonicity with a negligible restoring force. Moreover, one can expect that the respective distortion of the $Fd\bar{3}m$ structure being dynamic at high temperatures becomes static at lower ones. More direct evidence, however, has to come from high resolution diffraction experiments that should evidence on electronic redistribution within the $S_6$ octahedra or a large atomic displacement parameter of $Cd^{2+}$ in the fluctuation regime.

## IV. Conclusions

Summarizing, Raman scattering experiments on stoichiometric single crystals of $CdCr_2S_4$ prepared by a Br transport reactions show pronounced phonon anomalies that are evidence for a symmetry reduction and Cr off-centering in the cubic unit cell. The resulting enhanced electronic polarizability of displacements that modulate the Cr-S distance is proposed as a microscopic mechanism for the observed large magnetocapacitive effects.

We acknowledge support by DFG and FRSF of Ukraine Grant No.F29.1/014.

## *Reference List*

1. T. Kimura, et al., Nature (London) **426**, 55 (2003); N. Hur, et al., ibid. **429**, 392 (2004); Th. Lottermoser, et al., ibid. **430**, 541 (2004); T. Goto, et al., Phys. Rev. Lett. **92**, 257201 (2004); N. Hur, et al., ibid. **93**, 107207 (2004); B. B. Van Aken, et al., Nature Materials **3**, 164 (2004); Y. Tokura and Sh. Seki, Advanced materials **22**, 1554-1565 (2010).
2. H. Zheng, et al., Science **303**, 661 (2004); I. Levin, et al., Adv. Matter **18**, 2044 (2006); R. Ramesh and N. A. Spaldin, Nature Materials 6, 21 (2007).
3. M. Fiebig, J. Phys. D: Appl. Phys. **38**, R123 (2005).
4. N. A. Spaldin and M. Fiebig, Science **309**, 391 (2005); C. Ederer and N. A. Spaldin, Current Opinion in Solid State and Materials Science **9**, 128-139 (2005); N.A. Hill, J. Phys. Chem. B **104**, 6694-6709 (2000).
5. J. Hemberger, et al., Nature (London) **434**, 364 (2005).
6. H. Gobel, Journ. Mag. Magn. Mater. **3**, 143 (1976).
7. P. Lunkenheimer, et al., Phys. Rev. B **72**, 060103(R) (2005).
8. E. Dagotto, *Phase Separation and Collossal Magnetoresistance*, (Springer, Berlin) 2002.
9. G. Catalan, J. F. Scott, Nature **448**, E4 (2007); J. Hemberger, P. Lunkenheimer, R. Fichtl, H.–A. Krug von Nidda, V. Tsurkan, A. Loidl, Nature **448**, E5 (2007) and references therein.
10. T. Rudolf, et al., Phys. Rev. B **76,** 174307 (2007) and references therein.
11. H. Schmid and E. Ascher, J. Phys. C **7**, 2697 (1974) and references therein.
12. E. F. Steigmeier and G. Harbecke, Phys. Kondens. Materie **12**, 1, (1970).
13. N. Koshizuka, et al., Phys. Rev. B21, 1316 (1980) and references therein; N. Sanford, et al., Phys. Rev. Lett. **50**, 1803 (1983).
14. Intensity anomalies in Ref. 12 are spurious due to a strongly temperature dependent reference mode. The later study in Ref. 13 did not have sufficient resolution to allow a separate observation of splitted modes.
15. H. D. Lutz, et al., Z. Naturforsch. A **45**, 893 (1990) and references therein.
16. C. J. Fennie and K. M. Rabe, Phys. Rev. B **72**, 214123 (2005).
17. K. Wakamura, T. Arai, J. Appl. Phys. **63**, 5824 (1988).
18. IR modes at $T = 10$ K have been reported in Ref. 17 as LO modes at 105.2, 251.2, 352.1, 395.0 and as TO modes at 101.8, 248.8, 323.5 and 379.0 cm$^{-1}$. Almost the same values have been obtained in Ref.10 under extrapolation to T = 0 K.
19. G. Güntherodt and R. Zeyher, in *Light Scattering in Solids IV*, edited by G. Güntherodt and M. Cardona (Springer, Berlin, 1984) and references therein.
20. W. J. Miniscalco, G. McCollum, N. G. Stoffel, and G. Margaritondo, Phys. Rev. B **25**, 2947 (1982).
21. Fingerprints are higher order multiphonon scattering, broadening of modes and fluorescense effects, see W. Kauschke, et al., Phys. Rev. B **36**, 7469 (1987).
22. Implications of $F\bar{4}3m$ for spinel compounds can be found in Ref. 11.
23. J. Zwinscher, et al., Journ. Solid State Chem. **118**, 43 (1995).
24. M. Isobe, et al., Journ. Phys. Soc. Jap. **71**, 1848 (2002).
25. P. G. Radaelli, et al., Nature **416**, 155 (2002).
26. N. Shanthi, et al., J. Solid State Chem. **155**, 198 (2000).
27. N. Ray and U. V. Waghmare, Phys. Rev. B **77**, 134112 (2008).
28. R. G. Pearson, Proc. Nat. Acad. Sci. USA **72**, 2104 (1975).
29. J. K. Burdett and O. Eisenstein, Inorg. Chem. **31**, 1758 (1992).
30. F. Boucher, et al., Acta Cryst. B **51**, 952 (1995).
31. J. B. Neaton, et al., Phys. Rev. B **71**, 014113 (2005).
32. C. P. Sun, C. L. Huang, C. C. Lin, J. L. Her, C. J. Ho, J.-Y. Lin, H. Berger, and H. D. Yang, Appl. Phys. Lett. **96**, 122109 (2010) ; and PRB **79**, 214116 (2009).